\documentclass[11pt,twoside]{article}

\usepackage{asp2006-mod}
\usepackage{graphicx}

\begin{document}
\title{Trends for Outer Disk Profiles}   
\author{Peter Erwin, Michael Pohlen, Leonel Guti\'{e}rrez, and John E. Beckman}   
\affil{}    

\begin{abstract} 
The surface-brightness profiles of galaxy disks fall into three main classes,
based on whether they are simple exponentials (Type I), bend down at large
radii (Type II, ``truncations'') or bend up at large radii (Type III,
``antitruncations'').  Here, we discuss how the frequency of these different
profiles depends on Hubble type, environment, and the presence or absence of
bars; these trends may herald important new tests for disk formation models.
\end{abstract}


\section{The Diversity of Galaxy Disk Profiles}

Recent studies using moderately large samples of nearby disk galaxies have
demonstrated that ``exponential'' disks actually fall into three categories of
surface-brightness profiles \citep{erwin05,pohlen-trujillo,erwin08}.  For a
more detailed discussion of the background and context, we direct the reader
to those papers and to the contribution by Pohlen et al.\ (this volume).
Here, we briefly discuss some preliminary results relating disk profiles to
general galaxy properties, including Hubble type and degree of barredness, as
well as evidence for environmental dependence.

\section{Trends with Hubble Type and with Bars}

The left panel of Figure~1 shows the frequency of different profile types
along the Hubble sequence, using the galaxies from \citet{pohlen-trujillo},
\citet{erwin08}, and Guti\'{e}rrez et al.\ (2008, in prep).  For simplicity,
we group the profiles into ``truncations'' (Type II, including both
``classical'' truncations [CT] and Outer-Lindblad-Resonance [OLR] breaks) and
``non-truncations'' (Types I and III).  It is clear that truncations of
various types are most common in the \textit{latest} Hubble types.  This is
consistent with the reported high frequency of truncations from studies of
edge-on disks \citep[e.g.,][]{vdk82,kregel04}, since these studies have
concentrated on late-type spirals (principally Sc--Sd).

The right panel of Figure~1 shows the distribution of profiles types as a
function of bar strength, using the standard RC3 classifications, for
early-type disks (S0--Sb) in the field.  The frequency of Type II profiles is
clearly higher in barred galaxies; most of this is probably related to the
Outer Lindblad Resonance of the bars \citep{erwin07}.  We can also see that
the frequency of Type III profiles is \textit{anti}-correlated with bars (a
trend also present if we use numerical measures of bar strength).  This
suggests that whatever process produces antitruncations also weakens and
destroys bars, or else that bar formation tends to suppress this process; in
either case, this places useful constraints on scenarios of disk profile
formation \citep[e.g.,][]{elmegreen06,younger07}.

\begin{figure}
\centerline{\includegraphics[width=12cm]{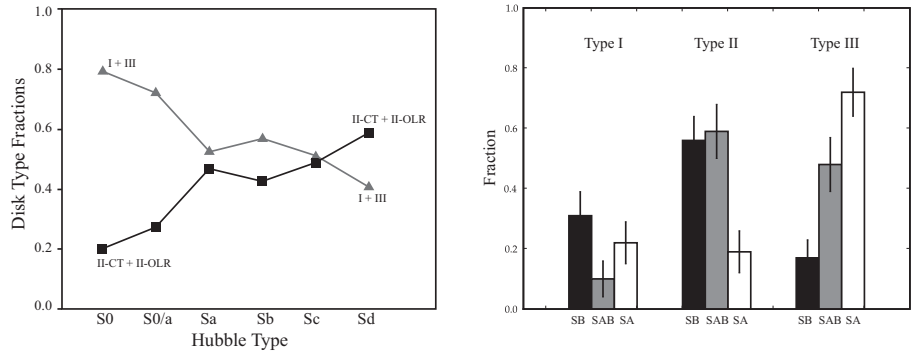}}
\caption{\textbf{Left:} General trends of disk-profile type with Hubble type.
We separate ``truncations'' (Type II profiles) from non-truncated profiles
(single-exponential Type I and ``antitruncated'' Type III; see Erwin et al.\
2008).  Note that the majority of Type II profiles in S0--Sb galaxies are
OLR-related, in contrast to the ``classical'' truncations that dominate
late-type spirals.  \textbf{Right:} The distribution of bar strengths (black =
SB, gray = SAB, white = SA) as a function of disk-profile type for early-type
(S0--Sb) field galaxies.  Type II profiles are clearly more common in barred
galaxies, but rare in unbarred galaxies; conversely, Type III profiles
(antitruncations) are least common in strongly barred galaxies and most
common when there is no bar.}
\end{figure}

\section{Disk Profiles and Galaxy Environment}

We are currently investigating whether outer disk profiles are affected by
their environment.  Preliminary results point to a dramatic difference in disk
profiles between the Virgo Cluster and the local field environment (including
galaxies in groups), at least for barred S0--Sb galaxies: about half the field
galaxies have Type II profiles, but only 10\% of the Virgo galaxies do.  This
suggests a strong role for the cluster environment in modifying outer disk
formation, something of potential relevance for, e.g., models of S0 formation.



\end{document}